\documentclass{revtex4}

\usepackage{epsfig}
\usepackage{amssymb}
\usepackage{graphicx}
\begin{document}
\title{ Quenching of antihydrogen gravitational states by surface charges }
\author{A.Yu. Voronin$^{1,2}$, E.A. Kupriyanova$^{1,2}$, A. Lambrecht$^3$, V.V. Nesvizhevsky$^4$, S. Reynaud$^3$.}
\affiliation{ $^1$ P.N. Lebedev Physical Institute, 53 Leninsky
prospect, 117924 Moscow, Russia.
\\
$^2$ Russian Quantum Center, 100 A, Novaya street, Skolkovo, 143025, Moscow, Russia.
\\
$^3$ Laboratoire Kastler Brossel, UPMC-Sorbonne Universit\'es, CNRS, ENS-PSL Research University, Coll\`ege de France, Campus Jussieu, 75252, Paris, France.%Gab
\\
$^4$ Institut Max von Laue - Paul Langevin (ILL), 71 avenue des Martyrs, F-38042, Grenoble, France.
%\\
%$^5$ Institut de Recherche sur les lois Fondamentales de l'Univers, CEA-Saclay, 91191, Gif-sur-Yvette, France.
%\\
%$^6$ Department of Quantum Chemistry, Uppsala University, Box 518,
%SE-75120 Uppsala, Sweden.
}

\begin{abstract}

We study an effect of quenching of antihydrogen quantum states near material surface in the gravitational field of the Earth by local charges randomly distributed along the mirror surface. The quenching mechanism reduces the quantum reflection probability because of additional atom-charge interaction and nonadiabatic transitions to excited gravitational states. Our approach is suitable for accounting for quenching caused by any kind of additional interaction with the characteristic range much smaller than the typical gravitational state wave-length.

\end{abstract}

\maketitle
\section{Introduction}
Discovery of gravitational states of ultra-cold neutrons (UCN), i.e. quantum states in the gravitational field of the Earth above a reflecting surface \cite{Nature1}, opened an interesting field of fundamental studies of quantum systems bound by gravity \cite{nesv00,nesv03,NVP,EPJC, NeutrWaveGuide,ResGranit}.
Attractive properties of such states are their extremely small bounding energy on one hand and mesoscopic spatial size on the other hand.
Such properties enable  searches for manifestations of nonstandard physics \cite{Axion07,JenkePRL14}  (including axion-like interactions, non-newtonian short range gravity, etc).

Extension of this approach to the case of gravitational properties of antihydrogen ($\bar{H}$) is possible due to the phenomenon of quantum (over-barrier) reflection from attractive but steep antiatom-surface potential \cite{FJM,Dufour2013qrefl,Dufour2013porous}. This phenomenon is responsible for existence of long-living quasi-stationary states of $\bar{H}$ in the gravitational field of the Earth above material surface \cite{voro05,voro05l,GravStates}. Study of gravitational quantum states of $\bar{H}$  opens perspectives for using precise spectroscopic and interferometric methods for measuring the gravitational mass of $\bar{H}$ \cite{Voro14resgrav,VoroNDLRKF14,Shaping} and testing Weak Equivalence Principle (WEP) with antiatoms \cite{AlphaGrav,Yam,Aegis1}.

Gravitational states of (anti)atoms could be much more sensitive to electric fields near material surface because of much larger polarizibility of atoms compared to neutrons.
Stray electric fields near material surface are known to be of significant importance for experiments with ultracold atoms.
There are several known sources of stray fields.
So called patch fields arise near polycrystalline metal surface because of different work fuctions of polycrystalline grains \cite{HerrNich, RzchHender, StrayFieldTheory,PatchEffect}.
They are affected by the surface contamination with adsorbed atoms or molecules producing dipole layers \cite{AdsAtomSurfElField, AdsorbElField}.
The intensity of patch fields near gold surfaces was measured by monitoring electric Stark shift in Rydberg atoms \cite{StrayFields} or using Kelvin Probe Force Microscopy \cite{KPFM}.
In case of the ground state of (anti)Hydrogen atom, the effect of such fields is weaker than the Casimir-Polder attraction.

However, $\bar{H}$ atoms bouncing on the surface could produce local charges on the surface due to annihilation in the bulk of the material mirror.
An excess of energy released in the annihilation process is capable to ionize numerous surrounding atoms.
This could be a possible source of accumulation of local charges on the surface.
Such residual charges on the surface could significantly modify mean interaction between an (anti)atom and a material  surface and quench gravitational states.
In this paper we study the effect of quenching of gravitational states as well as  transitions from low to highly excited gravitational states caused by the interaction with charges localized on the surface.

\section{Surface Charge-atom interaction}
In the following we will study the case of charges, localized on the surface within an area with a typical size $l_c$ small compared to the scattering length of $\bar{H}$ atom on van der Waals-Casimir-Polder potential $l_c\ll |a_{CP}| \sim 30 $ nm \cite{voro05,voro05l,GravStates}.
This in the region, where quantum reflection takes place (at distances of the order of $|a_{CP}|$).
The field produced by such charges can be treated as the field of a point-like charge $Q$.

The interaction between an atom and a point-like charge at such distances  is described by the polarization potential:
\begin{equation}
V_{pol}(z,\rho)=-\frac{\alpha_p Q^2}{2(z^2+\rho^2)^2}.
\end{equation}
Here $\alpha_p=9/2$ is the dipole polarizibility of (anti)Hydrogen atom, $z$ is the atom-surface distance, $\rho$ is the charge-atom distance in the surface plane.

The Schr\"{o}dinger equation, which describes $\bar{H}$ motion in the gravitational field of the Earth above a flat mirror with local charges on it, is:

\begin{equation}\label{ShrEq}
\left( -\frac{\hbar^2}{2m}\Delta+Mgz+V_{CP}(z) +V_{pol}(z,\rho)-E \right) \Phi(z,\vec{\rho})=0.
\end{equation}
Here $m$ and $M$ are $\bar{H}$ inertial and gravitational masses (which we distinguish in view of their comparison in future experiments), $V_{CP}(z)$ is van der Waals-Casimir-Polder potential, $E$ is the energy of $\bar{H}$.

This equation is supplied with the full absorption boundary condition at the surface ($z=0$), which can be formulated as the condition of absence of a reflected wave from the surface \cite{voro05}:
\begin{equation}\label{Boundary}
\Phi(z\rightarrow 0,\vec{\rho})=\frac{1}{\sqrt{p(z,\rho)}} \exp\left(-\frac{i}{\hbar} \int p(z',\rho) dz'\right)\chi(\vec{\rho}).
\end{equation}
Here $p(z,\rho)=\sqrt{2m(E-V_{CP}(z)-V_{pol}(z,\rho)-Mgz)}$  is the classical momentum of $\bar{H}$, and $\chi(\vec{\rho})$ is the wave-function of  planar motion.

Let us mention that such a boundary condition is valid due to the fact that WKB approximation holds with increasing accuracy when $z$ tends to zero. This boundary makes scattering insensitive to any details of interaction at the distances where WKB holds.

The reflection amplitude is determined by those parts of potential, where WKB is broken (so called badlands \cite{FJM, Df}) and where a reflected wave is generated \cite{voro05,voro05l}.
For pure van der Waals-Casimir-Polder potential these are distances of the order of $\sqrt{2m C_4}$, where $C_4$ is related to the asymptotic behaviour of $V_{CP}(z)=-c_4/z^4$ at large $z$.

The presence of $V_{pol}(z,\rho)$ not only modifies the properties of $\bar{H}$-surface interaction in the region, which determines quantum reflection, but also couples normal and tangential motions of $\bar{H}$.

In the following we will be interested in the lifetime and population of gravitational states during $\bar{H}$ motion along the mirror surface. It was shown in \cite{GravStates} that characteristic scales of van der Waals-Casimir-Polder interaction $l_{CP}= \sqrt{2m C_4}/\hbar=0.003$ $\mu$m are much smaller than the typical wave-length characteristic for gravitational interaction $l_g =\sqrt[3]{\frac{\hbar^{2}}{2mMg}}=5.871$ $\mu$m.

We will restrict our treatment to such surface charges that the scattering length on the potential $V_{pol}(z,0)$  characteristic scale is much smaller than the gravitational wave-length  $l_{pol}=\sqrt{m \alpha Q^2}/\hbar \ll l_g$, i.e. $Q \ll 10^4 e$.
In this case the shift and width of gravitational states due to the combined effect of short-range potentials $V_{CP}(z)$ and $V_{pol}(z,\rho)$ can be described via a complex scattering length $a(\rho)$ \cite{GravStates}, calculated for a fixed value of $\rho$:

\begin{eqnarray}\label{ScLength}
\left[ -\frac{\hbar^2}{2m}\Delta+Mgz-E \right] \Phi(z,\vec{\rho})=0,\\
\frac{d}{dz} \ln\left(\Phi(z\rightarrow 0,\vec{\rho})\right)=-\frac{1}{a(\rho)}.
\end{eqnarray}
The complex scattering length $a(\rho)$ can be evaluated from an asymptotic of the zero-energy solution $\Phi_0(z\rightarrow \infty,\vec{\rho})$ of the  Schr\"{o}dinger equation:
\begin{equation}\label{ShrEq0}
\left( -\frac{\hbar^2}{2m}\frac{d^2}{dz^2}+V_{CP}(z) +V_{pol}(z,\rho) \right) \varphi(z,\vec{\rho})=0,
\end{equation}
with the boundary condition (\ref{Boundary}). Here $\rho$ is treated as a parameter. The mentioned above asymptotic of such a solution has a form:
\[ \varphi (z\rightarrow \infty , \rho )=z-a( \rho ). \]

Thus the combined effect of van der Waals-Casimir-Polder and polarization potential is accounted by means of a $\rho$-dependent boundary condition.
In the following we will develop an approximation for solving above equations,  based on the classical treatment of planar motion of $\bar{H}$ atoms.

\section{ Time-dependent model}
In the following we will consider the planar motion (parallel to the surface) of $\bar{H}$ as classical motion along a straight  trajectory:
\begin{equation}
\vec{\rho}(t)=\vec{\rho}_0+\vec{v} t,
\end{equation}
where $\vec{\rho}_0$ is the initial planar position of the atom.

Equation  (\ref{ScLength}) turns into a time-dependent  Schr\"{o}dinger equation:
\begin{eqnarray}\label{TD1}
i\hbar\frac{d\Phi(z,t)}{dt}=\left[ -\frac{\hbar^2}{2m}\frac{d^2}{dz^2}+Mgz \right] \Phi(z,t),\\
\Phi(z\rightarrow 0,t)=z-a(t).
\end{eqnarray}
Time dependence of the complex scattering length is given through  time dependence of the planar charge-atom distance $\rho_i(t)$:
\begin{equation}\label{Rho}
\rho_i(t)=|\vec{\rho}_i^c-\vec{\rho}_0-\vec{v} t|,
\end{equation}
where $\vec{\rho}_i^c$ is position of $i$-th charge on the surface.

Equation system (\ref{TD1}) can be transformed into another form, more convenient for perturbation treatment.
The following complex coordinate and wave-function transformation:
\begin{eqnarray}
z'(t)=z-a(t),\\
\Phi(z,t)=\Psi(z',t)\exp\left(-i\frac{Mg}{\hbar} \int a(t) dt\right)
\end{eqnarray}
turns the equation system (\ref{TD1}) into the following equation system with a time-independent boundary condition:
\begin{eqnarray}\label{TD2}
i\hbar \frac{d\Psi(z',t)}{dt}=i\frac{d a(t)}{dt} \frac{d\Psi(z',t) }{dz'}+\left[ -\frac{\hbar^2}{2m}\frac{d^2}{dz'^2}+Mgz' \right] \Psi(z',t),\\
\Psi(z'\rightarrow 0,t)=0.
\end{eqnarray}

Expanding the time-dependent wave function in the basis of gravitational states and using the known expression for matrix elements of momentum operator between gravitational wave-functions we come to the following set of coupled equations for the time dependent amplitudes of gravitational states $C(t)$:
\begin{equation} \label{Ceq}
\frac{dC_i}{dt}=\frac{1}{l_g}\frac{d a(t)}{dt}\sum_k\frac{1}{\lambda_i-\lambda_k}\exp\left(-i \omega_{ki} t\right) C_k(t).
\end{equation}
Here $\omega_{ik}$ is the transition frequency between gravitational levels, $\lambda_i$ are the solutions of eigen-value problem for gravitational states:
\[\mathop{\rm Ai}(-\lambda)=0.\]

Here $\mathop{\rm Ai}(x)$ is the Airy-function \cite{abra72}.
In the following we will treat the case of a typical time of passage near the charge $\tau=\sqrt{ m \alpha_p Q^2}/(v\hbar)$  small compared to the typical gravitational time $\tau_g=\frac{\hbar}{\varepsilon_g}=0.001 $ s.
Perturbative solution  (\ref{Ceq}) for the transition amplitude with the initial condition $C_k(0)=\delta_{1k}$ in case of one surface charge is:
\begin{equation}\label{DiracSol}
C_k(t)=-\frac{iMg}{\hbar} \left(\int_{-\tau/2}^{\tau/2}a(t)\exp \left( -i\omega_{1k}t \right)dt-2a_{CP}\frac{\sin\left(\omega_{1k}\tau/2\right)}{\omega_{1k}}\right) \exp\left(-i\omega_{k1} t_0\right ).
\end{equation}
Here $t_0$ is the time instant when the atom approaches to a surface charge at a minimum planar distance.

In the limit $\omega_{1k}\tau \ll 1$ the above equation takes the form:
\begin{equation} \label{DiracSol0}
C_k(t)=-i\frac{Mg}{\hbar} \left(\int_{-\tau/2}^{\tau/2}a(t) dt-a_{CP}\tau\right) \exp\left(-i\omega_{1k} t_0\right ).
\end{equation}

In case of multiple charges, distributed on the surface and separated by large distances compared to $\sqrt{2 m \alpha_p Q^2}/\hbar$, one gets:
\begin{equation} \label{DiracSol1}
C_k(t)=-i \frac{Mg}{\hbar} \left( \int_{-\tau/2}^{\tau/2}a(t) dt-a_{CP}\tau\right) \sum_n\exp\left(-i\omega_{1k} t_n\right ).
\end{equation}
Here $t_n$ is the instant of the closest approach to $n$-th charge.

Corresponding transition probability within the perturbation theory approximation is:
\begin{equation}\label{probability}
P=\left|\frac{Mg}{\hbar} \left(\int_{-\tau/2}^{\tau/2}a(t)\exp \left( -i\omega_{1k}t \right) dt-2a_{CP}\frac{\sin\left(\omega_{1k}\tau/2\right)}{\omega_{1k}}\right)\sum_n\exp\left(-i\omega_{k1} t_n\right )\right|^2\exp\left(-2\frac{Mg}{\hbar} \int |\mathop{\rm Im} a(t)| dt\right).
\end{equation}
The probability of transition to low excited gravitational states in the above approximation is the same for any such state:
\begin{equation}\label{P0}
P=\left|\frac{Mg}{\hbar} \int_{-\infty}^{\infty}\left(a(t) -a_{CP}\right)dt \sum_n\exp\left(-i\omega_{1k} t_n\right )\right|^2\exp\left(-2\frac{Mg}{\hbar} \int |\mathop{\rm Im} a(t)| dt\right).
\end{equation}
We performed formal transition to infinite limits of integration, as  $a(t)\rightarrow a_{CP}$ for $|t|\gg \tau$. Indeed, such transformation is ensured by fast convergence of a corresponding integral for time  large compared to the typical time-of-passage near scattering center.

We will be interested in the probability of transition to all excited states, which can be obtained from (\ref{probability}) by summing over all states.
Substituting summation by integration over transition frequency and using the fact that $a(t)$ is an even function of $t$ we get the following expression for $P$:
\begin{equation}\label{Ptot}
P= \frac{M^2g^2 N \pi\tau_g}{ \hbar^2}\int_{-\infty}^{\infty}|a(t)-a_{CP}|^2 dt \exp\left(-2\frac{Mg}{\hbar} \int |\mathop{\rm Im} a(t)| dt\right).
\end{equation}
Here $N$ is the number of scattering centers along trajectory; we took into account that such centers are distributed stochastically; factor $\tau_g$ appears when summation over gravitational states is substituted by integration over transition frequency $dn\rightarrow d\omega \tau_g$.

It is instructive to compare expressions (\ref{Ptot}) and (\ref{P0}). One can see from the above comparison, that fast perturbation with a characteristic duration time $\tau$ effectively populates $n= \pi \tau_g/\tau$ gravitational states.

Let us mention that time dependence of the scattering length appears through time dependence of antiatom-scattering center planar distance (\ref{Rho}).
Introducing a minimal distance $d$ between scattering center and strait line planar trajectory of antiatom, we get for the transition probability  on one scattering center:

\begin{equation}\label{Pd}
P(d)\simeq\frac{M^2g^2  \pi\tau_g}{v \hbar^2}\int_{-\infty}^{\infty}|a(d^2+x^2)-a_{CP}|^2 dx .
\end{equation}
In the above expression we assume $\exp\left(-2\frac{Mg}{\hbar} \int |\mathop{\rm Im} a(t)| dt\right)\simeq 1$, keeping only first order terms in the transition probability.

It is useful to introduce an effective transition radius (an analog to the scattering cross-section in 3d case) according to the following expression:
\begin{equation}\label{Dtr}
d_{tr}=\int_{-\infty}^{\infty}P(\rho)d\rho,
\end{equation}
with $P(\rho)$ given by expression (\ref{Pd}).

With this definition, expression (\ref{Ptot}) can be reformulated as follows:
\begin{equation}
P_t\simeq d_{tr}L \sigma ,
\end{equation}
here $\sigma$ is the planar density of scattering centers, and $L$ is the characteristic length of a mirror.
Corresponding effective "transition" width  is given by the following expression:
\begin{equation}
\Gamma_{t}=\hbar d_{tr}\sigma v.
\end{equation}
Taking into account (\ref{Pd}) we get for $\Gamma_t$:
\begin{equation}\label{GammaT}
\Gamma_t=\frac{2M^2g^2 \sigma \pi^2\tau_g}{ \hbar}\int_{0}^{\infty}|a(\rho^2+x^2)-a_{CP}|^2 \rho d\rho.
\end{equation}

The decay  probability of (each) gravitational state within above approximations after passing one scattering center additional to constant decay rate on a plane surface is given by the following expression:
\begin{equation}\label{Pdec}
P_{in}(d)=1-\exp\left(\frac{2Mg}{\hbar v} \mathop {\rm Im} \int_{-\infty}^{\infty}(a(d^2+x^2)-a_{CP}) dx\right).
\end{equation}
Taking into account smallness of this probability we get:
\begin{equation}\label{PdecS}
P_{in}(d)\simeq \frac{2Mg}{\hbar v} \left |\mathop {\rm Im} \int_{-\infty}^{\infty}(a(d^2+x^2)-a_{CP}) dx \right |.
\end{equation}

Introducing effective decay radius according to the following expression:
\begin{equation}\label{Din}
d_{in}=\int_{-\infty}^{\infty}P_{in}(\rho)  d\rho,
\end{equation}
we get the following expression for the decay probability during the flight along the mirror:
\begin{equation}
R=1-\exp\left(-d_{in}L \sigma -2Mg |\mathop{\rm Im} a_{CP}|\frac{L}{\hbar v}\right).
\end{equation}
Effective decay width is given by the following expression:
\begin{equation}
\Gamma_{in}=\hbar d_{in}\sigma v+ 2Mg|\mathop{\rm Im} a_{CP}|.
\end{equation}
Taking into account (\ref{PdecS}) we get for $\Gamma_{in}=\Gamma_{d}+\Gamma_{CP}$:
\begin{eqnarray}\label{Gammad}
\Gamma_{d}=4Mg \pi \sigma \left |\mathop {\rm Im}\int_{0}^{\infty}(a(\rho^2)-a_{CP}) \rho d\rho \right |,\\
\Gamma_{CP}=2Mg|\mathop{\rm Im} a_{CP}|\label{GammaCP}.
\end{eqnarray}

As one can see, the quenching mechanism of an initial gravitational state is due to the additional inelastic width, which antiatom gains while passing through the region with a modified atom-surface potential, as well as due to transitions to other gravitational states.
However, two quenching mechanisms have different orders of magnitude.
It follows from (\ref{Dtr}) and (\ref{Din}) that the ratio of corresponding quenching rates is given by:
\begin{equation}\label{ratio}
\frac{d_{tr}}{d_{in}}=\frac{\pi\int_0^{\infty}\left|a(\rho^2)-a_{CP}\right|^2 \rho d\rho }{2 l_g \left |\mathop {\rm Im}\int_{0}^{\infty}(a(\rho^2)-a_{CP}) \rho d\rho \right |}.
\end{equation}
The above expression is less than unit in the range of $Q$-values, where our approximations are valid.

The dynamics of such quenching is determined by time dependent scattering length $a(t)$ on the sum of van der Waals-Casimir-Polder and polarization potentials.
In the following sections we will calculate $a(t)$ and evaluate the decay width and transition probabilities of the ground state.

\section{ Complex scattering length and numerical results}

\begin{figure}
  \centering
\includegraphics[width=100mm]{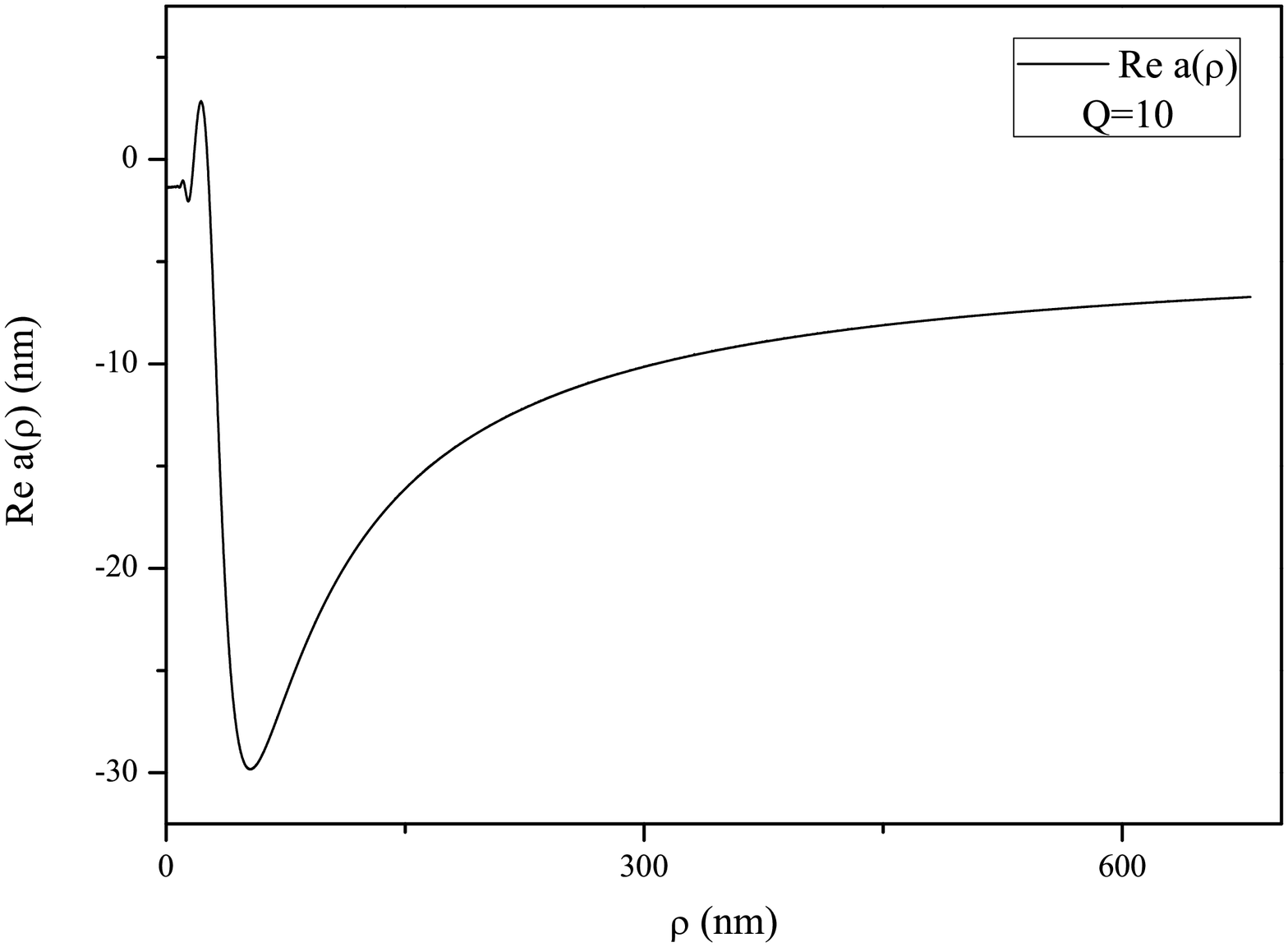} % alexei
%\centerline{\psfig{file=ReflCasimir2.eps,width=12.0cm}} %hahrevised mode2
\caption{Real part of the scattering length $\mathop{\rm Re} a(\rho)$ as a function of the planar antiatom-charge distance. $Q=10 e$.}\label{Rea10}
\end{figure}

\begin{figure}
  \centering
\includegraphics[width=100mm]{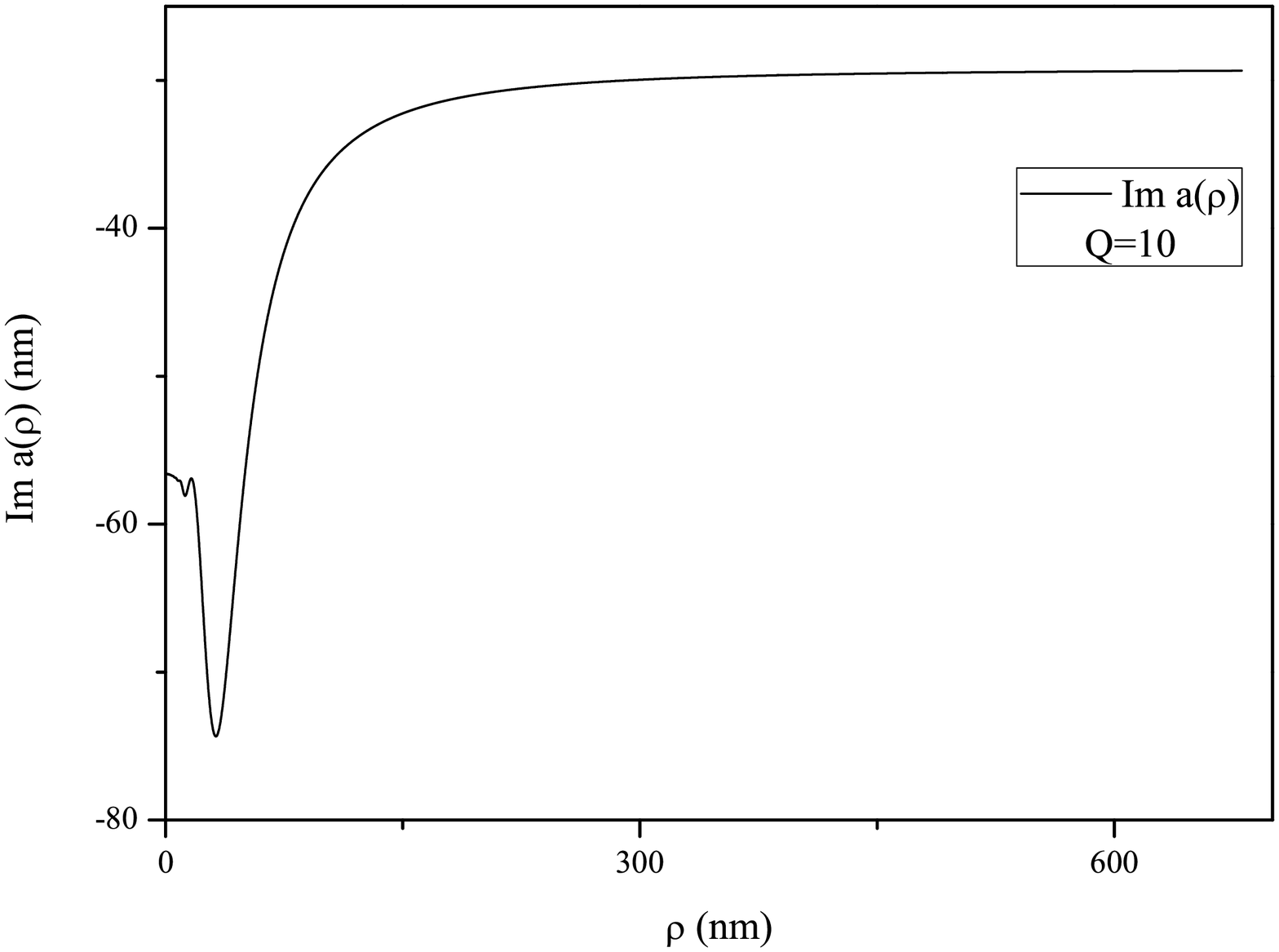} % alexei
%\centerline{\psfig{file=ReflCasimir2.eps,width=12.0cm}} %hahrevised mode2
\caption{Imaginary part of the scattering length $\mathop{\rm Im} a(\rho)$ as a function of the planar antiatom-charge distance. $Q=10 e$.}\label{Ima10}
\end{figure}

\begin{figure}
  \centering
\includegraphics[width=100mm]{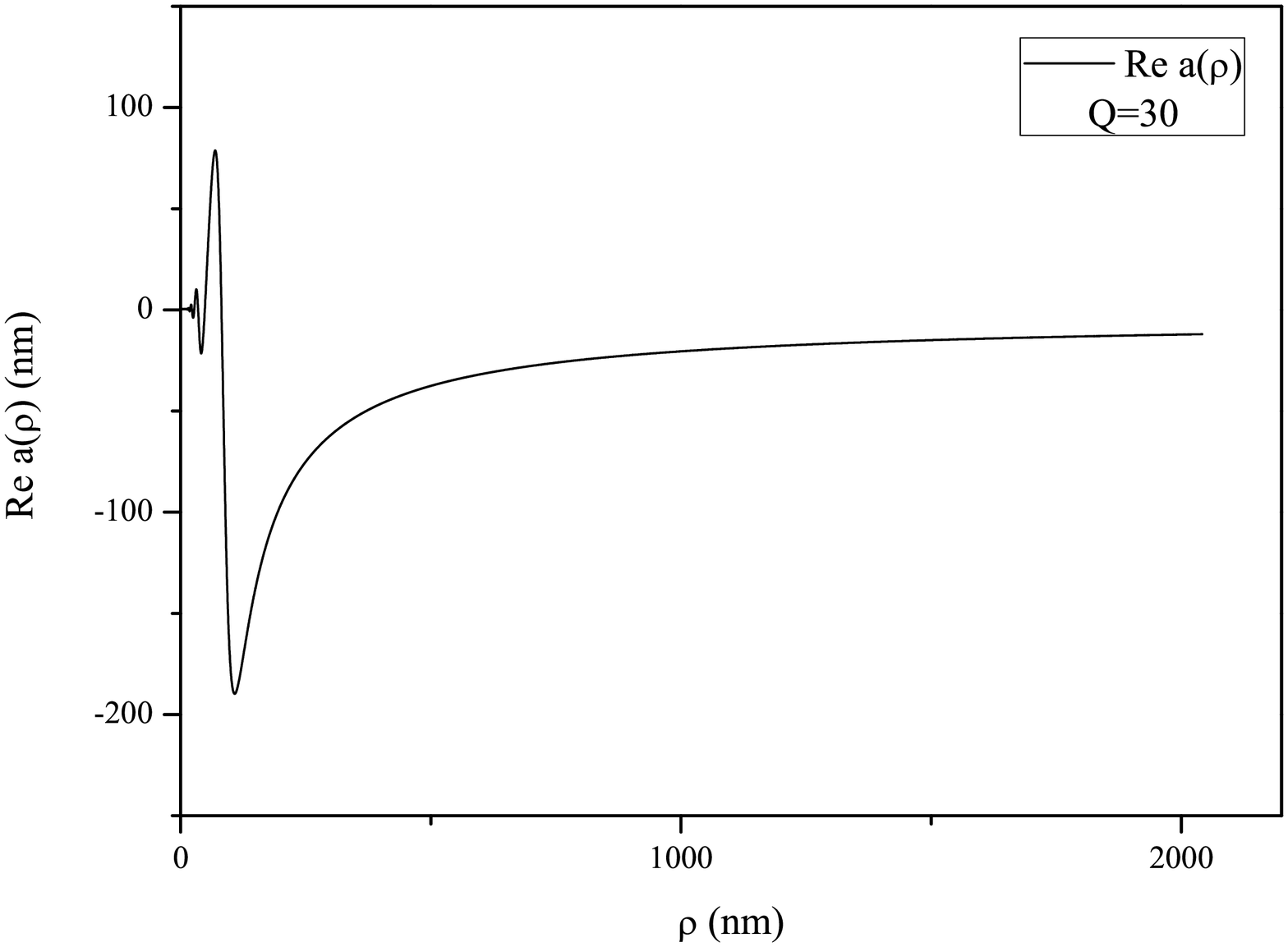} % alexei
%\centerline{\psfig{file=ReflCasimir2.eps,width=12.0cm}} %hahrevised mode2
\caption{Real part of the scattering length $\mathop{\rm Re} a(\rho)$ as a function of the planar antiatom-charge distance. $Q=30 e$.}\label{Rea30}
\end{figure}

\begin{figure}
  \centering
\includegraphics[width=100mm]{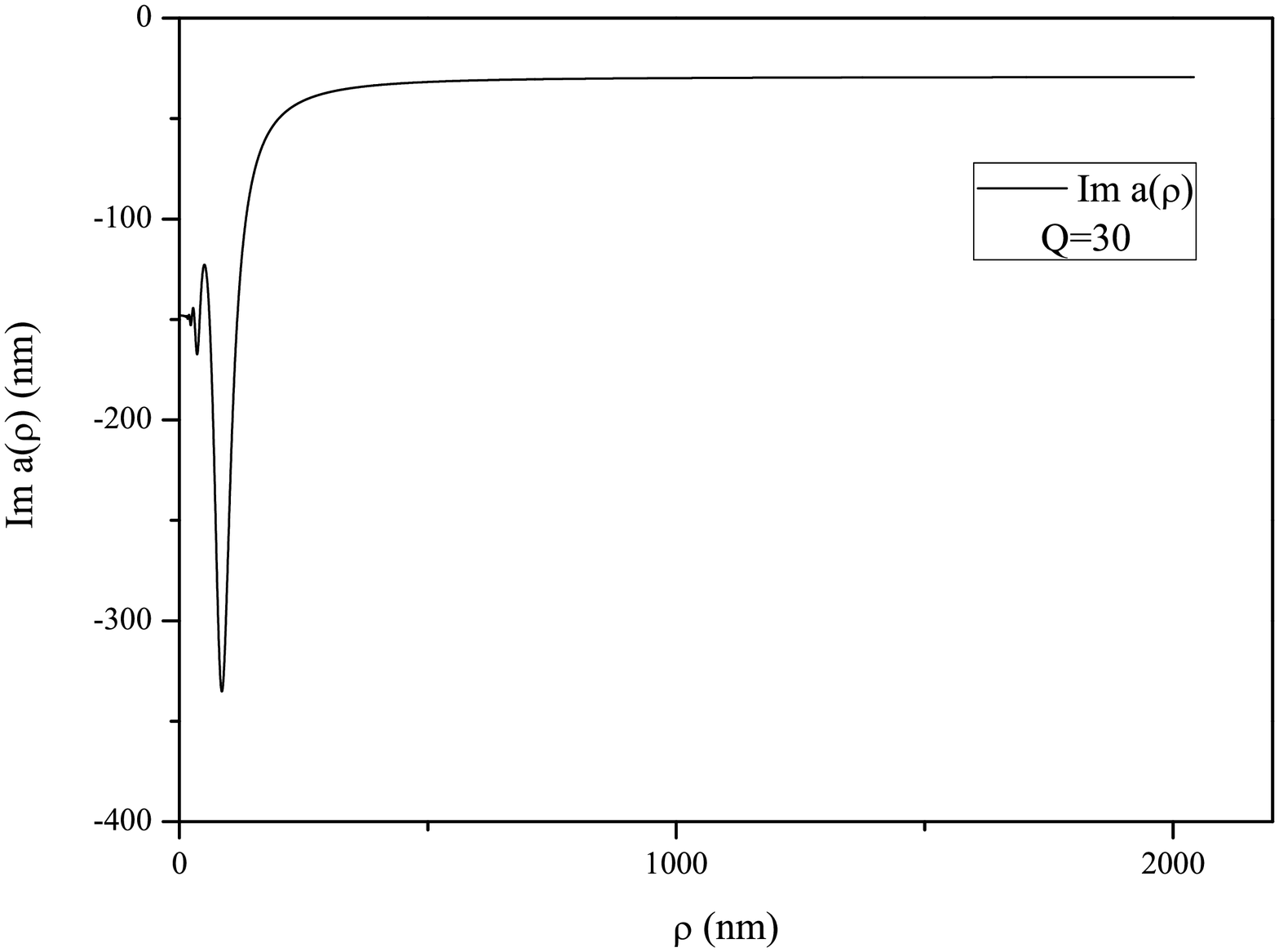} % alexei
%\centerline{\psfig{file=ReflCasimir2.eps,width=12.0cm}} %hahrevised mode2
\caption{Imaginary part of the scattering length $\mathop{\rm Im} a(\rho)$ as a function of the planar antiatom-charge distance. $Q=30 e$.}\label{Ima30}
\end{figure}

The complex scattering length is evaluated by numerically solving equation (\ref{ShrEq0}) with boundary condition (\ref{Boundary}).
Real and imaginary parts of the scattering length $\mathop{\rm Re} a(\rho)$, $\mathop{\rm Im} a(\rho)$, as a function of the planar antiatom-scattering center distance $\rho$, are shown in Fig. \ref{Rea10}, Fig. \ref{Ima10}, Fig. \ref{Rea30}, Fig. \ref{Ima30}.
%%%%%%%%%%%%%%%%%%%%%%%%%%%%%%%%%%%%%%%%%%%%%%%%%%%%%%%%%%%%%%%%%%%%%%%%

%%%%%%%%%%%%%%%%%%%%%%%%%%%%%%%%%%%%%%%%%%%%%%%%%%%%%%%%%%%%%%%%%%%%%%%%

%%%%%%%%%%%%%%%%%%%%%%%%%%%%%%%%%%%%%%%%%%%%%%%%%%%%%%%%%%%%%%%%%%%%%%%%%%%
%%%%%%%%%%%%%%%%%%%%%%%%%%%%%%%%%%%%%%%%%%%%%%%%%%%%%%%%%%%%%%%%%%%%%%%%

%%%%%%%%%%%%%%%%%%%%%%%%%%%%%%%%%%%%%%%%%%%%%%%%%%%%%%%%%%%%%%%%%%%%%%%%%%%
%%%%%%%%%%%%%%%%%%%%%%%%%%%%%%%%%%%%%%%%%%%%%%%%%%%%%%%%%%%%%%%%%%%%%%%%

%%%%%%%%%%%%%%%%%%%%%%%%%%%%%%%%%%%%%%%%%%%%%%%%%%%%%%%%%%%%%%%%%%%%%%%%%%%
%%%%%%%%%%%%%%%%%%%%%%%%%%%%%%%%%%%%%%%%%%%%%%%%%%%%%%%%%%%%%%%%%%%%%%%%%%%

It is worth mentioning rapid oscillations of $\mathop{\rm Re} a(\rho)$ and sharp maxima of $|\mathop{\rm Im} a(\rho)|$. The physical reason for such oscillations in brief consists of the fact that a superposition of van der Waals-Casimir-Polder potential $V_{CP}(z)$ and polarization potential $-\frac{\alpha_p Q^2}{2(z^2+\rho^2)^2}$ produces two regions of over-barrier reflection.
The separation between such regions depends on $\rho$ (for given $Q$).
Such over-barrier reflection from two separated regions supports a near-threshold over-barrier resonance as soon as the distance between reflection regions matches a sort of standing wave condition.
Appearance of a near-threshold resonance is reflected in a fast change of the scattering length as a function of $\rho$.

\begin{figure}
  \centering
\includegraphics[width=100mm]{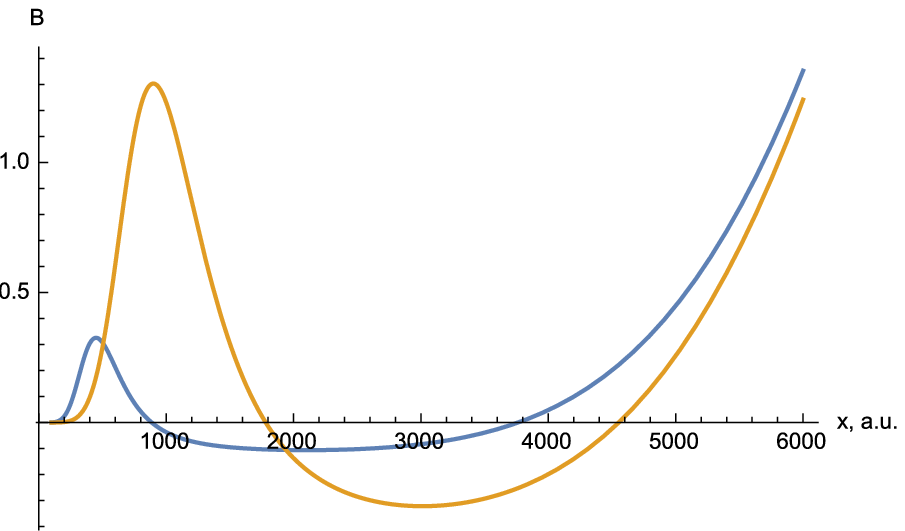} % alexei
%\centerline{\psfig{file=ReflCasimir2.eps,width=12.0cm}} %hahrevised mode2
\caption{Badland functions for two values of $\rho=1000$ a.u. (blue line) and $\rho=2000$ a.u. (yellow line), $Q=30$, energy $E$ is taken equal to the energy of the ground gravitational state.}\label{badland}
\end{figure}

In order to demonstrate this statement we study the so called "badland" function:
\begin{eqnarray}
B(z)=\hbar^2 \left(\frac{p''(z)}{2p^3(z)}-\frac{3}{4}\left(\frac{p'(z)}{p^2(z)}\right)^2 \right),\\
p(z)=\sqrt{2m\left(E-V_{CP}(z)-V_{pol}(z)\right )}.
\end{eqnarray}
In the regions where $|B(z)|\geq 1 $ WKB approximation fails and quantum reflection takes place.
In Fig. \ref{badland} we show $B(z)$ for $\rho=1000$ and $\rho=2000$ a.u. and $Q=30$.
One can see that for $\rho=2000$ a.u. there is a region of WKB validity between two badland regions, which ensures "standing wave" resonance condition.
%%%%%%%%%%%%%%%%%%%%%%%%%%%%%%%%%%%%%%%%%%%%%%%%%%%%%%%%%%%%%%%%%%%%%%%%

\begin{figure}[h]
  \centering
\includegraphics[width=100mm]{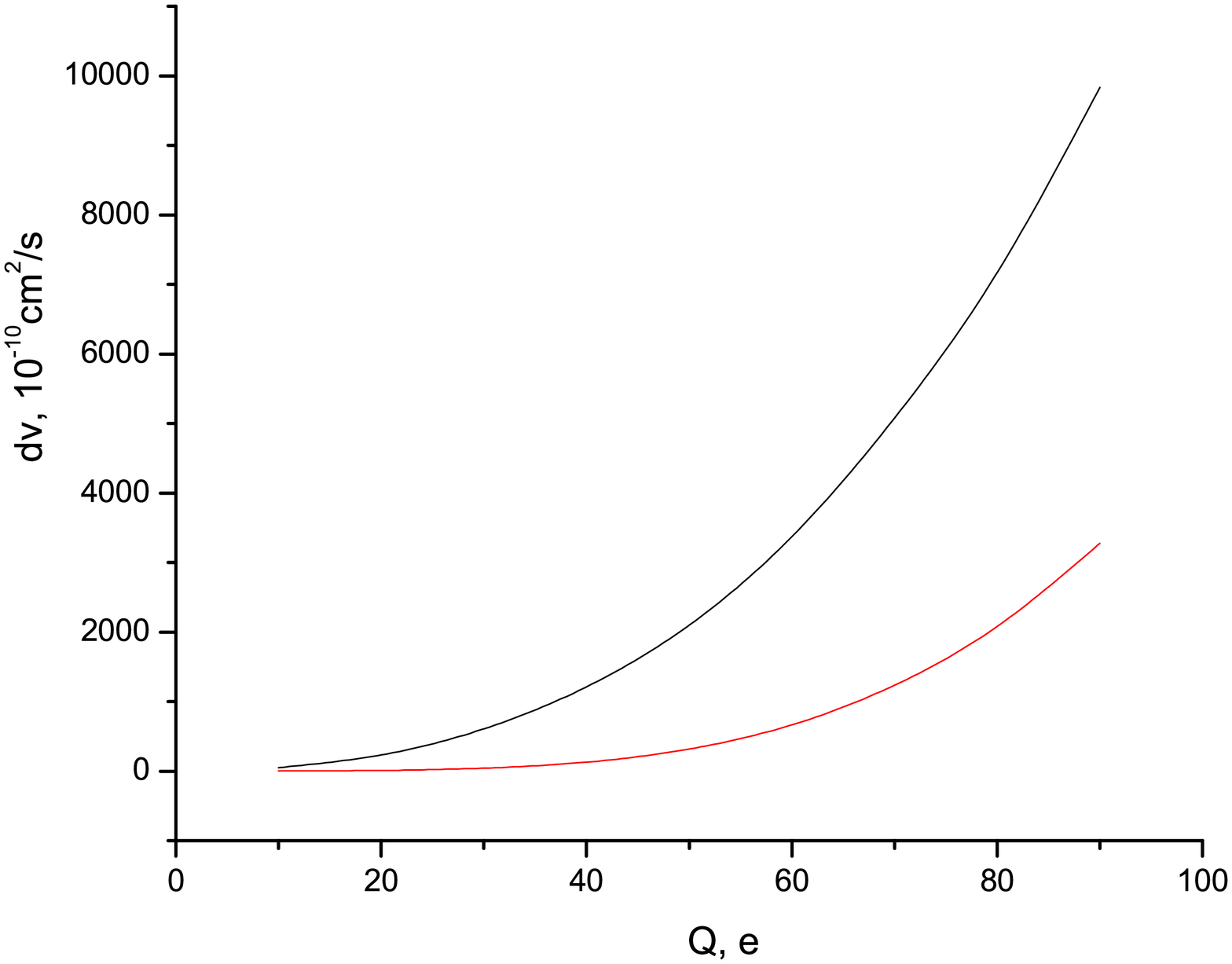} % alexei
%\centerline{\psfig{file=ReflCasimir2.eps,width=12.0cm}} %hahrevised mode2
\caption{Black line  $d_{in}v$, red line- $d_{tr}v$, shown  as a function of scattering center charge $Q$. }\label{dvplot}
\end{figure}

%%%%%%%%%%%%%%%%%%%%%%%%%%%%%%%%%%%%%%%%%%%%%%%%%%%%%%%%%%%%%%%%%%%%%%%%%%%
The value of the product $d_{in}v$ and $d_{tr}v$ is given in Fig. \ref{dvplot}.
In particular,  $d_{in}=6.06\cdot 10^{-12}$ m and $d_{tr}=4.0 \cdot10^{-13}$ m for $v=1$ $ m/s$ and charge $Q=30$.
The effective scattering decay width (\ref{Gammad}) becomes comparable with the decay width on plane surface (\ref{GammaCP}) when the surface density of scattering centers $\sigma_c \simeq 10^{12}$ $ m^{-2}$ equally charged with $Q=30$.
Such a critical density $\sigma_c\simeq 10^{11}$ $m^{-2}$ is an order of magnitude smaller for the charge of scattering centers $Q=80$.
%%%%%%%%%%%%%%%%%%%%%%%%%%%%%%%%%%%%%%%%%%%%%%%%%%%%%%%%%%%%%%%%%%%%%%%%

%%%%%%%%%%%%%%%%%%%%%%%%%%%%%%%%%%%%%%%%%%%%%%%%%%%%%%%%%%%%%%%%%%%%%%%%

\section{Conclusion}

We studied the quench mechanism for $\bar{H}$ gravitational states above material surface caused by local electric charges, stochastically distributed along the mirror surface.
We performed our analysis for the case of the characteristic time of charge-antiatom interaction smaller than the characteristic time typical for the gravitational interaction.
This approximation enables us to obtain closed form results for the decay probability due to annihilation on the surface as well as due to transitions to excited states.
Such results are expressed in terms of the scattering length on the superposition of van der Waals-Casimir-Polder and polarization potentials.
Let us note that the developed formalism is not specific for quenching by surface charges only, but also can be applied to quenching by any additional interaction, when the characteristic spatial radius is smaller than the  gravitational wave-length, so that the scattering length approximation is justified.
In particular, the obtained results could be applied to quenching by patch effects or surface roughness on the plane mirror.

The leading quenching mechanism in case of individual surface charges smaller than $100$ $e$ consists of enhanced annihilation of $\bar{H}$ in the surface.
This enhancement of annihilation originates from the decrease of quantum reflection on one hand and due to specific nearthreshold resonances, which appear in the superposition of van der Waals-Casimir-Polder and polarization potentials on the other hand.

For individual charges on the surface equal to $30$ $e$ with a density of surface scattering centers equal to $10^{12}$ $ m^{-2}$, the corresponding enhanced decay width would become comparable with the one on the plane surface without charges. This shows that the effect of local charges on lifetime of gravitational states can be neglected as long as the numbers for surface charges remain smaller than the values discussed above. As such densities of local charges can be experimentally controlled, false effects due to charges on the surface can be eliminated.

\section {Acknowledgement}

One of the authors (AV) is grateful to N. Kolachevsky for postulating the problem and fruitful discussions.

\end{document}